\documentclass[aps,prl,twocolumn,showpacs,superscriptaddress,groupedaddress]{revtex4-1}  
\usepackage{graphicx}  
\usepackage{dcolumn}   
\usepackage{bm}        
\usepackage{amssymb}   
\usepackage{url}

\begin{document}

\preprint{FI-UPM-CCS/1}

\title{Quantum Simulation of the Factorization Problem}
\thanks{This paper is dedicated to Professor Emeritus Jos\'{e} Luis S\'{a}nchez-G\'{o}mez.  }%

\author{Jose Luis Rosales}
 \email{Jose.Rosales@fi.upm.es}
\author{Vicente Martin}
 \email{Vicente@fi.upm.es}
\affiliation{%
 Center for Computational Simulation, \\
 ETS Ingenieros Inform\'{a}ticos, Universidad Polit\'{e}cnica de Madrid,\\
Campus Montegancedo, E28660 Madrid. 
}

\date{\today}

\begin{abstract}
Feynman's prescription for a quantum simulator was to find a hamitonian for a system that could serve as a computer. P\'olya and Hilbert conjecture was to demonstrate Riemann's hypothesis through the spectral decomposition of hermitian operators. Here we study the problem of decomposing  a number  into its prime factors, $N=xy$, using such a simulator. First, we derive the hamiltonian of the physical system that simulate a new  arithmetic function, formulated for the factorization problem, that represents the energy of the computer. This function rests alone on the primes below $\sqrt N$. We exactly solve the spectrum of the quantum system without resorting to any external ad-hoc conditions, also showing that it obtains, for $x\ll \sqrt{N}$, a prediction of the prime counting function that is almost identical to Riemann's $R(x)$ function. It has no counterpart in analytic number theory and its derivation is a consequence of the quantum theory of the simulator alone.

\end{abstract}

\pacs{03.67.Ac, 03.67.Lx, 02.10.De, 89.20.Ff}
\keywords{Quantum Computing, Quantum simulators, Number Theory, Prime Numbers, Factoring, RSA }
\maketitle

The computational complexity assumption~\cite{sieves} to find  the prime factors of a large number $N$ is the basis for the security of the ubiquitous RSA, a cornerstone of the public key cryptosystems so widely used in our digital society.
  However, despite the many mathematical and computational advances, the classical complexity of the factorization problem is still unknown. Fortunately, the best classical algorithms known scale worse than polynomially in the number of bits of $N$: By now, the building blocks of the cyberinfrastructure still resist.

Nonetheless, in the quantum world, factoring is an easy problem that requires only polynomial resources using Shor's algorithm\cite{shor}. This amazing result raises new questions about the relationship between quantum mechanics and number theory and, more generally, with physics; a connection dating back to  P\'olya and Hilbert~\cite{Montgomery, Schumayer}, who laid a program to prove Riemann's hypothesis through the spectrum of physical operators.   
However, the physical realization of Shor's algorithm is still limited to proof of concept demonstrations, far away from factoring numbers of the size used in real-world cryptosystems.

An alternative would be to build the solutions in Hilbert space of a quantum simulator performing factorization, instead of going through the route of a gate-based, fully programmable, quantum computer. The key idea, following the pioneering suggestions of Feynman~\cite{Feynman}, is to translate factoring arithmetics into the  physics of a  device whose superposition of  states  mimics the problem i.e.: a factoring (analog) computer. The states of the simulator would be the solutions of some hermitian operator depending only on the number that we want to factorize. 
Moreover, by simply using the computer over different values of $N$, a quantum factoring simulator must be capable to access the statistics of the prime numbers.  Thus, it might provide insight on fundamental problems in number theory following the P\'olya and Hilbert program.   
 Here we propose a new approach to  the factorization problem based on the physics of a bounded  hamiltonian  that corresponds to a new arithmetic function defined for this problem. The values of this new function should correspond, in the quantum theory, to eigenvalues of the simulator. To the best of our knowledge, this is the first example of a quantum system whose spectrum supports the P\'olya and Hilbert conjecture.

First, to bind the hamiltonian, we need a problem definition leading to a finite Hilbert space. For this we define a factorization ensemble for a given $N$~\cite{supplemental}. Suppose that we want to factorize $N$. A simple trial division algorithm will require to inspect all the primes $x$ less or equal than $\sqrt{N}$, i.e., a total of $\pi(\sqrt{N})$ trials will be required. The factorization ensemble of $N$ is defined as  the set of all pairs of primes that, when multiplied, give numbers $N_k$ with the property $\pi(\sqrt{N_k})=j$, where $j=\pi(\sqrt{N})$.

The solution to the factorization problem consists then in finding the appropriate pair in the factorization ensemble, that we will denote as $\mathcal{F}(j)$.

Then, to build a bridge between number theory and quantum mechanics, we   redefine the factorization problem introducing a single-valued arithmetic function, computed for a pair of primes $(x_k, y_k)$  in the ensemble of $N$. After, we transform this function into a hamiltonian mapping the arithmetics of factorization to the physics of a classical system; finally we obtain the quantum observable (operator) corresponding to the energies of the classical counterpart. Thus, obtaining the factor of $N$ is equivalent to measuring the energy of this simulator.

The cardinality of the factorization ensemble is thus important, since, given this interpretation, it is the dimension of the Hilbert space  associated to the observable. It can be derived~\cite{supplemental} as a corollary of Theorem 437 in \cite{FA} for the special case of the product of two primes.

\begin{equation}
\label{eq:Fcardinality}
|\mathcal{F}(j)|\simeq  \sqrt{N}(\log\log \sqrt{N}+o(1))\sim \sum_{x_{k}=2}^{\sqrt{N}}\frac{\sqrt{N}}{x_{k}}.
\end{equation}
where the sum is taken over the primes.
$\;$ Moreover, given these estimates, one would expect  $\sqrt{N}/x_{k}$ as the number of possible different co-primes $y_{k}$ per each $x_{k}$.  The new arithmetic function  that represents the hamiltonian of the system, should be symmetric in the factors of $N$ and also include an explicit dependence on $j$.

A simple function with these properties  for $N_{k}=x_{k}y_{k}\in \mathcal{F}(j)$, is:
\begin{equation}
\label{eq:E}
	E(x_{k},y_{k})=\frac{\pi(x_{k})\pi(y_{k})}{j^2}, 
\end{equation}
where $\pi(x)$ is the prime counting function. Note that, knowing an exact rational value of $E$ there necesarily  exists   a single solution  of the  equation $E=\pi(x)\pi(N/x)/j^2$ in the ensemble. Obviously $E=E(x)$.   Moreover the behavior of $E(x)$, similarly to $\pi(x)$, has two components, a regular plus a oscillatory  one~\cite{supplemental}:

  \[E(x) = 1+\epsilon(N,x),\] 
  where   $\epsilon(N,x) = u(N,x)^2 + \epsilon_{fl}(N,x)$.

Here, $\epsilon_{fl}(N,x)$, the oscillating function,  depends on the zeros of Riemann's $\zeta$~\cite{PNT}, while $u$ is a regular function that can be approximated for $N>>1$ as 
\begin{equation}
\label{eq:u}
u(N,x) =\gamma\log(\sqrt{N}/x),
\end{equation}
where  $\gamma= j/\sqrt{N}\sim 1/\log (\sqrt{N})$.

Let us introduce now two new arithmetic functions:
\begin{eqnarray}
p=\frac{\pi(y)-\pi(x)}{2 j}, \:
	q=\frac{\pi(y)+\pi(x)}{2 j}.
\end{eqnarray}
Of course, these are related to $E$, because after Euclid's factorization theorem there exist a single free parameter for the problem of factoring $N$ (the factor $x$ or, as we have reformulated here, the value $E$) 
\begin{equation}
\label{eq:iHO}
-p^2+q^2=E.
\end{equation}
Which has the form of  the energy of  the classical inverted harmonic oscillator whose trajectories can be parameterized as $q=E^{1/2}\cosh(t)$. From this point of view, along with the computation of $E$ from Eq.~\ref{eq:iHO}, we might also  consider variations in $p$ and $q$ due entirely to changes in $t$ at constant $E$. For large $N$, $t$ can be considered a quasi-continuum parameter  and it has indeed the meaning of the time variable in Hamilton's equations (i.e., $E$ is an adiabatic invariant of the variation)
\begin{equation}
\delta p=-\partial_q H\delta t,\:
\delta q=\partial_p H \delta t.
\end{equation}
$H$ being the hamiltonian on the canonical coordinates $p$ and $q$.
\begin{equation}	
	H(p,q)=\frac{1}{2}(-p^2+q^2).
\end{equation}
Moreover $p=\partial_q S(q)$, in terms of the Hamilton's principal function (the action) $S(q)$, obtaining the Hamilton-Jacobi equation
\begin{equation}
\label{eq:HJcond}
H(\partial_q S(q),q)=E/2;
\end{equation}
Eq.~\ref{eq:HJcond} is relevant because $q$ must be bounded in $\mathcal{F}(j)$ and therefore its solutions are confined trajectories in parametric space.
\begin{equation}
\label{eq:qmN}
\sqrt{E}\leq q\leq\frac{\pi(N/x_{m})+\pi(x_{m})}{2 j}=q_{m}, 
\end{equation}
for some $x_{m}$ in $\mathcal{F}(j)$. 

Now, the Hamilton-Jacobi constraint for $S(q)$ and quantum transformation theory allow us to 
obtain the momentum  operator acting on the wave functional $\psi_{E}(q)$ for  the q-numbers. $p\rightarrow-i\partial_q$; the hamiltonian constraint in Eq.~\ref{eq:iHO} becoming  a hermitian operator in our  coordinates acting on $\psi$. 
It is interesting to note that the same hamiltonian has been used previously, although through a different canonical transformation,  in the study of the distribution of Riemann's zeros~\cite{B-K}.

Hence,  Eq.~\ref{eq:iHO} transforms into
\begin{equation}
\label{eq:schro}
\psi_E(q)^{''}+q^2\psi_E(q)=E \psi_E(q),
\end{equation}
our coordinate space satisfies  $E^{1/2}\leq q \leq q_{m}$, and our quantum conditions should be
\begin{eqnarray}
\label{eq:constraints}
\psi_E(E^{1/2})=0 ,\:
\psi_E(q_{m}(N))= 0.
\end{eqnarray}

The Schr\"{o}dinger Eq.~\ref{eq:schro} and the Sturm-Liouville conditions  in Eq.~\ref{eq:constraints} define the eigenvalue problem leading to the quantization of $E$. It is important to note here that 
we do not have to impose any ad-hoc constraints to the wave function in order to reach the limits required for quantization. Now, a coordinate transformation $\rho=q^2$ and $\psi_E=R_E(\rho)\rho^{3/4}$, gives
\begin{equation}
\label{eq:coulscat}
R_E^{''}+\frac{2}{\rho}R_E^{'}-\frac{l(l+1)}{\rho^2}R_E+2\mu(r^2-\frac{z^2}{\rho})R_E=0,
\end{equation}
where $l=-1/4$, $\mu=1/2$, $r=1/2$ and $z^2=E/4$, transforms our equation in the tridimensional Schr\"{o}dinger equation for the coulombian scattering of two identical charged particles in their center of mass.

The general solution of Eq.~\ref{eq:coulscat} is 

\begin{eqnarray}
\label{eq:gensol}
R_E(\rho)=\rho^{-1/4}\Re\{ e^{-i\rho/2}[U(\alpha(E),3/2,i\rho)\\ \nonumber
+D_{0} F(\alpha(E),3/2,i\rho)]\}.
\end{eqnarray}

$F(a,b,c)$ and $U(a,b,c)$ are the confluent hypergeometric functions, $\alpha(E)=-i\frac{E}{4}+\frac{3}{4}$ and $D_{0}$ is a function of $E$ obtained from $R_E(E)=0$. After Eq.~\ref{eq:constraints}, the solution exists  if and only if the energy $E$ is real and exactly satisfies the quantum condition~\cite{supplemental}:
\begin{eqnarray}
\label{eq:quantization}
\Re\{\frac{F(\alpha,3/2,i\rho_{m})U(\alpha,3/2,iE)}{
F(\alpha,3/2,i E)U(\alpha,3/2,i\rho_{m})}\}=1;
\end{eqnarray}

Note that inverting Eq.~\ref{eq:quantization} provides an algorithm to get $x|N$ from $E$, the eigenvalue corresponding to the quantum stationary state of the simulator.

The hypothesis of the existence of the quantum simulator will be true if  and only if the spectrum of the simulator provides the statistics of the prime numbers.  

The problem requires the theory of scattering of nuclear charged particles \cite{lan}. Asymptotically, for $\rho\gg 1$, Eq.\ref{eq:coulscat} gives
\begin{equation}
\label{eq:asymp}
R_E \sim 1/\rho\;\sin(\rho/2-\frac{E}{4}\log\rho+\delta_{C}+\frac{7\pi}{8}+\delta_{0}).
\end{equation}
Here $\delta_{C}=\arg{\Gamma(\alpha)}$,  is  a shift in the distorted Coulomb wave for the asymptote and $\delta_{0}$ is obtained from the asymptotic formulas of $U(\alpha,3/2,i\rho)$ and $F(\alpha,3/2,i\rho)$ as  
\begin{equation}
\label{eq:cotdelta}
D_{0} e^{3\pi E/8}\cot\delta_{0}\rightarrow 1.
\end{equation}

Recall now that, in the ensemble, $E$ attains its maximum  at $\pi(3)=2$,
\begin{equation}
\label{eq:E3}
 \max{E}= 2\frac{\pi(N/3)}{j^2} \sim 1/3\gamma^{-1} =o(\log \sqrt{N}),
\end{equation}
it means that, for small prime factor candidates $x|N$, the values of $e^{3E\pi/8}$ in Eq.~\ref{eq:cotdelta} are $O(\sqrt{N})$ when we  expand $E$ in a series near $\frac{1}{3}\log \sqrt{N}$. This gets~\cite{supplemental}
 
  \[\delta_{0}=A\:\sqrt{N}\!\log E-h+\frac{\pi}{2},\] where $A$ and $h$  depend only on $N$.

From the asymptote in Eq.~\ref{eq:asymp}, the second quantum condition at $\rho_{m}=q_{m}^2$ \ imposes $R_E(\rho_{m})=0$. Therefore 
\begin{equation}
\label{eq:2qcond}
\delta_{C}+\delta_{
0}+\rho_{m}/2-E/4\log\rho_{m}+\frac{7\pi}{8}=n\pi,
\end{equation}
where $n$ is an integer number.
Redefine  $n=\lfloor \rho_{m}/(2\pi)\rfloor-k$, for some integer $k$ , $1\leq k\leq |\mathcal{F}(j)|$ (the convention taken that large $k$'s map the region $E\gg 1$). When $N\gg 1$, the leading term in Eq.~\ref{eq:2qcond} is precisely $\delta_{0}$, it yields to 
\begin{equation}
\label{eq:Avalue1}
 -A\frac{\sqrt{N}}{\pi}\log E +o(1/E^2)\rightarrow(k-\frac{h-\pi/2}{\pi}).
\end{equation}
Now we have  $E(\max k)=\max{E}$. Using  Eq.~\ref{eq:Fcardinality} with $\max{k} = |\mathcal{F}(j)| $ gets

\[A\rightarrow -\pi, \: h=O(\sqrt{N}),\] 
and $\delta_{0}\rightarrow -\pi \sqrt{N}\log E+ O(\sqrt{N})$; $h (N)$, contributes to the wave function as a global phase and can be fixed with a new re-definition of $n$ as previously done. 

Thus, from Eq.~\ref{eq:Avalue1} one obtains the solution $E(k)$  for $k\sim O(|\mathcal{F}(j)|)$, i.e., small prime factor candidates
\begin{eqnarray}
\label{eq:spectrum}
E\rightarrow C\gamma^{-\kappa}
\end{eqnarray}
where, for convenience, we  defined the variable $\kappa\equiv k/|\mathcal{F}(j)|$, and $C$ is a parameter depending on $N$.

It is possible to obtain a better insight of the meaning of Eq.~\ref{eq:spectrum} by transforming its dependence on the variable $\kappa$ on other in $u(N,x)$. This is possible because there is only one pair of co-primes in $\mathcal{F}(j)$ such that $N=xy$, implying  a relation $\kappa\rightarrow x$. To explore this, we can use a simple interpolating polynomial of degree 2 in two known primes, say $x=2$ and $x=3$, using the statistics of the primes in $\mathcal{F}(j)$: 

\begin{equation}
\label{eq:u_parabolic}
u(N,x) \simeq \alpha_{1}(N)\kappa -\alpha_{2}(N)\kappa^2.
\end{equation}
Eq. \ref{eq:u_parabolic} also satisfies that $u(N, \sqrt{N})=0$ at $\kappa=0$, forcing the constant term to be zero.
The result for $\mathcal{F}(304)$ is shown in Fig.~\ref{ukappa}.

\begin{figure}[hbtp]
\centering
\includegraphics[scale=0.55]{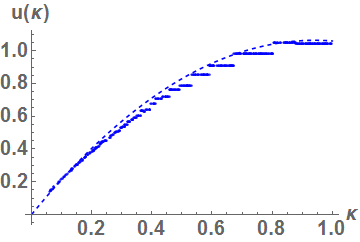}
\caption{ The function $u$ is represented versus $\kappa=k/|\mathcal{F}|$ for $x$ in the  factorization  ensemble with $j=304$.  The dashed line is the Lagrange polynomial in $x=2$ and $x=3$ ($u=2.26 \kappa-1.20\kappa^2$), although other pairs work the same. Remarkably the number of divisor candidates in $\mathcal{F}(j)$ for $x=2$, $x=3$ etc., satisfies empirically  the statistics predicted from our asymptotic estimates in Eq. \ref{eq:Fcardinality}.}
\label{ukappa}
\end{figure}

This solution, valid for any $N$, can also  be used as a  theoretical test of the quantum simulator. Let us  check explicitly that the statistics of the states corresponds to that of the primes. Simply inverting Eq. \ref{eq:u_parabolic} we get~\cite{supplemental}:
\begin{equation}
\label{eq:egamma}
E(x)\rightarrow C\gamma^{-\kappa(x)}.
\end{equation}

Now, directly from Eq.~\ref{eq:E}  and  recalling that asymptotically  $\pi(N/x)\rightarrow j/(1+u)\frac{\sqrt{N}}{x}$, we finally obtain~\cite{supplemental} for $x\ll \sqrt{N}$

\begin{equation}
\label{eq:piapprox}
\pi(x|N)\rightarrow \gamma x(1+u)E(x).
\end{equation}
For $x$ a prime candidate to factor $N$. This can be interpreted as a parametric family of curves enveloping $\pi(x)$.  Thus, we can determine the constant $C$ by simply matching some known value of $\pi(x)$ to the asymptote above. 

Further proof of the exactness of the results obtained here is to show that 
the expression of $\pi(x|N)$ in Eq.~\ref{eq:piapprox} actually does not depend on $N$, as can be deduced classically from the universality of the primes and that for $N\rightarrow \infty$ every prime should be in $\mathcal{F}(j)$. We have experimentally tested this in many cases. This is a necessary condition, but comes as an striking verification, since all the results arise from a purely quantum theory.

\begin{figure}[hbtp]
\centering
\includegraphics[scale=0.24]{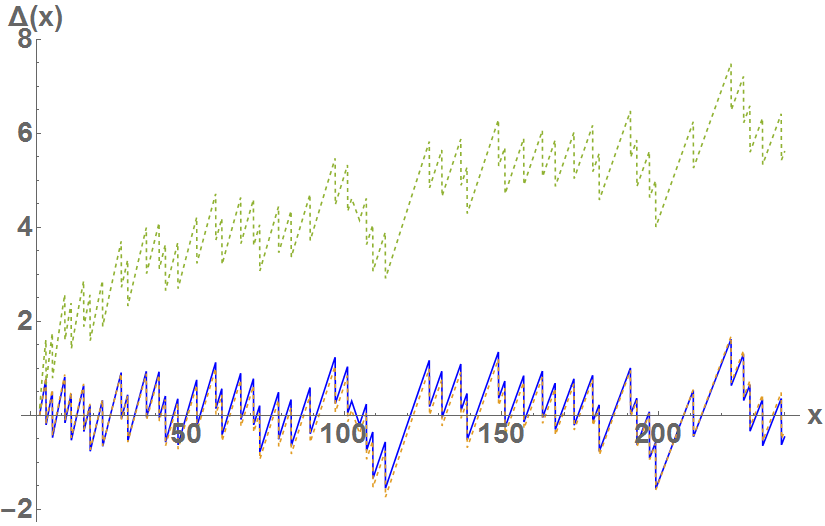}
\caption{ The functions $\Delta(x)=\pi(x|N)-\pi(x)$ calculated here (blue),  $R(x)-\pi(x)$ (dashed orange) and  $Li(x)-\pi(x)$ (dashed green) for $x$ in the  factorization ensemble with $j=3155$. Note how $\pi(x|N)$ fits perfectly to the best  analytical approximation given by $R(x)$ for $x\ll \sqrt{N}$.} 
\label{deltax}
\end{figure}

As  seen in Fig. \ref{deltax}, Eq. \ref{eq:piapprox} (for $x\ll \sqrt{N}$) is tantamount to the best possible approximation, given by Riemann function.  The result fully confirms the consistency of the quantum solution.

Eq.\ref{eq:u_parabolic} is just an element required for the calculations; it was obtained specifically to match, using the simplest possible polynomial, the function $u(N,x)$ in terms of the statistics of the primes in $\mathcal{F}(j)$. Note that $\kappa(x)$ must exists -- independently of our approximations -- and, according to the distribution of prime factor cadidates in the ensemble, should be a stepwise function.

To summarize, we introduced  new  concepts and  arithmetic functions that could play a significant role in the quantum factorization problem. The  Factorization Ensemble is the main one; it allows us to bind the hamiltonian of a quantum factoring simulator. Then, we have reformulated the factorization problem to that of finding a new parameter of the problem: the arithmetic function $E$; it corresponds to the energy eigenvalues of the simulator.  We show that the spectrum of the simulator gives 
in the semiclassical  quantization regime -- large $k$, i.e., $x\ll\sqrt{N}$-- the statistics of the primes. The compelling exactitude of this prediction justifies that both, the simulator and the new algorithm of factorization outlined, that inverts the quantum conditions (Eq.~\ref{eq:quantization})  for the coprime factor  $x=f(E)$ will work. The next step will be to find out a suitable physical system, described by this hamiltonian, to which the boundary conditions can be applied. The spectrum of the system will provide the E values that, through the inverse of the quantum conditions found in this paper, will finally give the factors.

As a final remark, this work supports indirectly P\'olya and Hilbert program\cite{Montgomery} to prove Riemann's hypothesis: the spectrum of the imaginary part of the zeros of  $\zeta(\sigma)$  should be eigenvalues of an hermitian operator. This being true,  it will imply, according to Riemann, the statistics of the primes $\pi(x)$. Here we evaluated $E(x)$ -- an eigenvalue of an hermitian operator-- obtaining an approximation to $\pi(x)$ for the primes in $\mathcal{F}(j)$. It suggests that,  perhaps, the truth of Riemann's hypothesis could   be found with the help of the functions and the approach introduced in this work, particularly- let's speculate with the physics of the hypothesis~\cite{Schumayer}- if the contributions of $\delta_C(E)$, for $E\sim 1$  to the spectrum of the energies of the simulator were correlated with those obtained for the arithmetic function $E$, computed from the zeros of  $\zeta(\sigma)$ on the critical line.

This work has been partially supported by Comunidad 
Aut\'{o}noma de Madrid, Project Quantum Information 
Technologies Madrid (QUITEMAD+), Project
No. S2013/ICE-2801 and by the Spanish Ministry of 
Economy and Competitiveness, project CVQuCo, Project 
No. TEC2015-70406-R. We thank Jes\'{u}s Mart\'{i}nez-Mateo for suggestions and  F. A. G. Lahoz for  pointing us to Ref. [6].


\section{ Supplemental material for "Quantum Simulation of the Factorization Problem" }


\maketitle

Given the interdisciplinary nature of the work, that straddles the apparently disparate fields of number theory and quantum physics, we are adding this Supplementary material. Its intention is to detail several of the new concepts introduced, give examples, make explicit some calculations and provide additional support to the reasoning and results in the manuscript.

\section{Factorization Ensemble} 

Suppose that we want to factorize a given number $N$. A simple trial division algorithm will require to inspect all the primes $x$ less or equal than $\sqrt{N}$, i.e., a total of $\pi(\sqrt{N})$ trials will be required. The factorization ensemble of $N$ is defined then as  the set of all pairs of primes that, when multiplied, give numbers $N_k$ with the property $\pi(\sqrt{N_k})=j$, where $j=\pi(\sqrt{N})$. 
\begin{equation}
\mathcal{F}(j)=\{ N_k= x_{k}y_{k}, : \pi(\sqrt{N_k})=j\},
\end{equation}
where $x_k$ and $y_{k}$ are primes.

All these numbers are equivalent when applying the prime trial division algorithm: they require the same number of operations and an hypothetical simulator will need the same resources to find the factors. On the other hand $\mathcal{F}(j)$ defines  a convenient neighbourhood of the number $N$  that we intend to factorize since, from the prime number theorem, asymptotically, $N_k\sim (\sqrt{N}-\nu_k \log \sqrt{N})^2$,  where the variable $\nu_k$ runs from $0$ to $O(1)$. The factorization ensemble will be composed by those $N_k$ that are the product of two primes and belong to this vicinity of $N$.
 
 For instance, let $N= 2003^2=4012009$, then, $j=304$. Yet, other possible $N_k$ in the ensemble are $N_1= 4021993=1019\cdot 3947$ and $N_2= 4026629=1291\cdot 3119$. Since $\pi(\sqrt{N})=\pi({N_1}^{1/2})=\pi({N_2}^{1/2})=304$ and, since they all are product of two primes, $N$, $N_1$ and $N_2$ are in the $j=304$ factorization ensemble $\mathcal{F}(304)$.
 
Thus,  the solution to the factorization problem consists  in finding the appropriate pair of primes in the factorization ensemble of $N$. 

Note that in case that the factors are not prime numbers, there is no single solution. Take for example:

 \begin{eqnarray*}
 \pi(541 \times 1223) \pi(1987) = 16107900 \\
 \pi(541 \times 1987) \pi(1223) = 16780000
 \end{eqnarray*}
  But, in any case, they do not belong to the Factorization ensemble, since all its elements are prime numbers.
  
  In the case that all of them are prime numbers, we might question if there exists prime factors $x_1, x_2$ and $x_3$ and $y_1, y_2$ such that:
  
   \begin{eqnarray*}
 \frac{\pi(x_1 x_2) \pi(x_3)}{j^2} = \frac{\pi(y_1) \pi(y_2)}{j^2} 
 \end{eqnarray*}
    
    The answer is that they might exist, but then  $ x_1, x_2 $ and $x_3$ does not belong to the factorization ensemble $\mathcal{F}(j)$, that is just composed by pairs of primes. (Note that to correctly pose the equation above we have to assume that $j=\pi(\sqrt N_1) = \pi(\sqrt(N_2))$ where $N_1= x_1 x_2 x_3$ and $N_2= y_1 y_2$). 
 
The total number of products, say $n$, of two primes less or equal than a given $N$ is, for $x_i$ prime (using the notation in \cite{FA}):
\begin{equation}
\pi_2(N)=\sum_{x_i\leq \sqrt{N}} \pi(\frac{N}{x_i})\sim \frac{N }{\log N}(\log\log N+O(1)).
\end{equation}
Now  the cardinality of $\mathcal{F}(j)$  can be calculated as the difference of $\pi_2(N)$ and the amount of those other products of two primes  $n\leq N^{'}\simeq (\sqrt{N}-\log \sqrt{N})^2$:

\begin{equation}
|\mathcal{F}(j)|\sim \pi_2(N)-\pi_2(N^{'}).
\end{equation}
This gives Eq. 1 in the main paper:
\begin{equation}
\label{eq:Fcardinality}
|\mathcal{F}|\sim  \sqrt{N}(\log\log \sqrt{N}+O(1))\sim \sum_{p}^{\sqrt{N}}\frac{\sqrt{N}}{p},
\end{equation}
and we  have used  the prime number theorem to compute the asymptotic value recalling the approximation\cite{INGHAM} 
\[
\sum_{p}^{\sqrt{N}} 1/p\sim\log\log \sqrt{N} + O(1).
\] 

Remark: In all the calculations we can take, for   $N_k$ in the ensemble, $\sqrt{N_k}\simeq \sqrt{N}$ whenever $N\gg 1$, since the diference $ \sqrt{N_k}- \sqrt{N}= O(\log \sqrt{N}$).

We can check the predicted cardinality of the ensemble for $j=304$:  
\begin{eqnarray*}
|\mathcal{F}(304)|\sim  \sqrt{N}(\log\log \sqrt{N}+O(1)) \\ \nonumber \simeq 4012009^{1/2}(\log\log 4012009^{1/2}+ 1)\simeq 6082
\end{eqnarray*}
versus the exactly computed $|\mathcal{F}|(304)=5760$, which is correct within the order of magnitude.

\section{Statistics.}

By simply inspecting Eq.\ref{eq:Fcardinality} it is directly deduced that, statistically,  one would expect  $\sqrt{N}/x_k$ as many possible different co-primes $y_{k}$ per each $x_{k}$  in the ensemble. This quantity  counts the cases $y_k=N_k/x_k$  in the factorization ensemble of the known $N$ we want to factorize.

\begin{figure}[hbtp]
\centering
\includegraphics[scale=0.34]{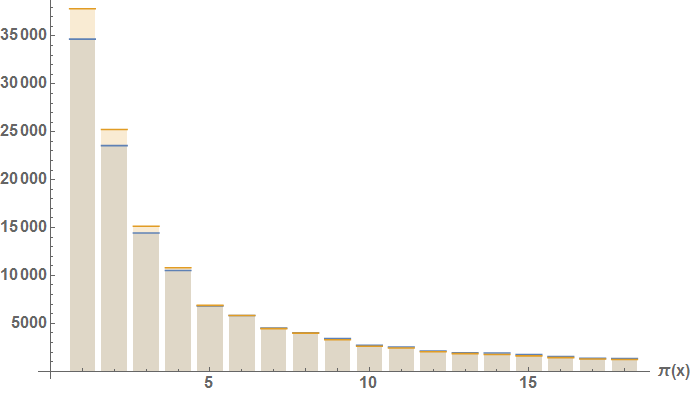}
\caption{ Statistics of the factorization ensemble for $j=7460$. The actual number of coprimes $y_k$ in the factorization ensemble $\mathcal{F}(7460)$  is represented (orange) as a function of $\pi(x_k)$ for the corresponding $x_k$.  The approximation from Eq. \ref{eq:Fcardinality}, uses $\sqrt{N}/x$ ($N=5731852681=75709^2$) and is plotted in blue to show its accuracy.}
\label{statistics}
\end{figure}

To clarify more this point, let's study again the statistics of $\mathcal{F}(j)$ with  $j=304$ example, considering possible coprimes pairs, $(x_k,y_k)$, in the ensemble $\mathcal{F}(304)$. To begin with, let's first take  $x=2$. Then one gets 1123 possible coprimes  from 	$y_1=2006021$,	$N_1= 4012042$ to $y_{1123}=2022043$, $N_{1123}=4044086$. Next prime is $x=3$, then,  $y_k$ in the ensemble goes through $y_{1124}=1337359$, $N_{1124}=4012077$ to $y_{1887}=1348033$,	$N_{1887}=4044099$. These figures fit perfectly well to the theoretical estimates because the ratio of possible coprimes for $x=2$ versus those corresponding to $x=3$ given by  $1123/764\simeq 1.47$, which is very similar to the statistically predicted value $3/2$. Of course, the larger the value of $j$, the more exact the statistical approximation becomes. To illustrate the accuracy of this point, we show in  Fig.\ref{statistics}, the actual  $j=7460$ statistics versus the theoretically derived cases for coprimes $y$ in the ensemble, $\sqrt{N}/x_k$.

\section{The spectrum $E_k$}

In the search for a classical counterpart of a possible quantum observable, we define for $N_{k}=x_{k}y_{k}\in \mathcal{F}(j)$ the arithmetic function:

\begin{equation}
\label{eq:E1}
	E_k=\frac{\pi(x_{k})\pi(y_{k})}{j^2}.
\end{equation}

The prime number theorem obtains  $E_k$ as a sum of a regular part plus an oscillating part $\epsilon_{fl}$  for large $N_k$: 

 \[E_k = R(x_k)R(y_k)/j^2+\epsilon_{fl}.\] 
 
Where $R(x)$ is the Riemann function and $\epsilon_{fl}$ depends basically on the zeros of Riemann's $\zeta$\cite{PNT} . 

The oscillatory part comes from the expansion 

  \[\pi(x)=R(x)-\sum_{\rho}{R(x^\rho)};\] 
where $\rho$ denotes all zeros (trivial and non-trivial) of the Riemann's $\zeta$ function.

The function  $\epsilon_{fl}$ is:

 $$
 \epsilon_{fl}(x,y)  = \frac{-1}{j^2} (f(x) R(y) +f(y) R(x) -f(x) f(y) ) 
 $$
 where $f(x) = \sum_{\rho} R(x^{\rho})$. The sum is, again, over all zeros of the Riemann's $\zeta$ function.

The regular part has an asymptotic quadratic behavior -- recall that, for $x\sim o(\sqrt{N})\gg 1$ one also has $R(x)\sim Li(x)\sim x/\log(x)$ --.
\begin{equation}
\label{eq:E}
R(x_k)R(y_k)/j^2 \sim 1+u(N_k,x)^2,
\end{equation}
where,
\begin{equation}
u(N_k,x) =\gamma_k\log(\sqrt{N_k}/x).
\end{equation}
Here, for convenience, we introduced the notation  $\gamma_k= j/\sqrt{N_k}$ (the difference between $\gamma_k$ and $\gamma = j/{\sqrt N}$ is negligible so we can simply take $\gamma_k\approx \gamma$ in all the calculations).

\begin{figure}[hbtp]
\centering
\includegraphics[scale=0.6]{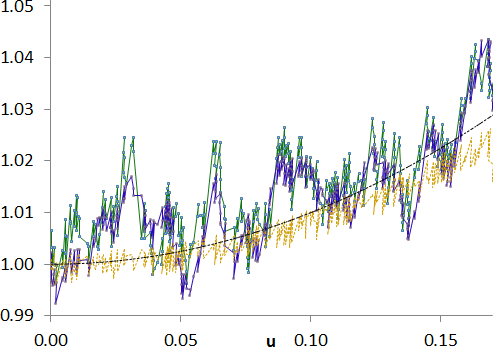}
\caption{ Plot of $E_k$ (in green) for $j=304$ ($N=2003^2=4012009$),  versus $u (N_k,x_k)$ in the ensemble. The region where $x\sim o(\sqrt{N})$, implying that $E\sim o(1)$ is represented. It  is seen that  $E= R(x_k)R(y_k)/j^2+\epsilon_{fl}$ is  a fluctuating function. $\epsilon_{fl}\sim O(N^{-1/4})$ depends essentially  on the sum of contributions from the Riemann's $\zeta$ zeros.  The blue line was computed using the sum of the contributions from the first $30$ Riemann's $\zeta$ zeros plus the regular part. The dotted black line is $E\simeq 1+ u(N,x_k)^2$. The orange line is  the regular function computed exactly for the numbers $N_k$ in the ensemble: $E\simeq 1+ u(N_k, x_k)^2$.}
\label{E}
\end{figure}

Again, to make this point more clear, we use our $j=304$ example. The exact  "spectrum" of $E_k$ is shown in Fig. \ref{E} as a function of $u(N_k,x_k)$.

\section{ Obtaining the quantum conditions from the  Schr\"{o}dinger equation.}

\subsection{Exact quantum condition.}
The general solution of the Schr\"{o}dinger equation for the simulator is (Eq.~13 in the main text):

\begin{eqnarray}
\label{eq:gensol}
R_E(\rho)=\rho^{-1/4}\Re\{ e^{-i\rho/2}[U(\alpha(E),3/2,i\rho)\\ \nonumber
+D_{0} F(\alpha(E),3/2,i\rho)]\}.
\end{eqnarray}
Now, in order to derive the quantum condition we need to recall that, after the given definition of $D_0$ and the Sturm-Liouville problem (Eq.~11 in the main text.),  one obtains from the condition $R(\rho_m)=0$
\begin{equation}
D_0= -U(\alpha(E),3/2,i\rho_m)/F(\alpha(E),3/2,i\rho_m),
\end{equation}
On the other hand, after the first Sturm-Liouville condition  one has at $\rho=E$: 
\begin{equation}
D_0= -U(\alpha(E),3/2,iE)/F(\alpha(E),3/2,iE).
\end{equation}
Finally, from these two equations we get the quantum condition (compatibility of the Sturm-Liouville problem):
\begin{eqnarray}
\label{eq:quantization}
\frac{F(\alpha,3/2,i\rho_{m})U(\alpha,3/2,iE)}{
F(\alpha,3/2,i E)U(\alpha,3/2,i\rho_{m})}=1;
\end{eqnarray}

Or, since $E$ is real, the more restrictive condition:
\begin{eqnarray}
\label{eq:quantization}
\Im\{\frac{F(\alpha,3/2,i\rho_{m})U(\alpha,3/2,iE)}{
F(\alpha,3/2,i E)U(\alpha,3/2,i\rho_{m})}\}=0;
\end{eqnarray} 

\subsection{Semiclassical quantization: asymptotic behavior of $\delta_0$.}

Recall that for large $\rho$, the asymptotic behavior of the Hypergeometric functions solutions of the Schr\"{o}dinger equation  requires that
\begin{equation}
\label{eq:cotdelta}
\tan \delta_{0}\rightarrow e^{3\pi E/8}D_{0}\sim \sqrt{N} D_0(E).
\end{equation}
The asymptotic behavior is valid for $E\sim \log \sqrt{N}$. Now, from its definition after the Sturm-Liouville condition at $\rho=E$,  
\begin{eqnarray*}
D_0(E)\sim f(E)\Re\{ -(iE)^{\alpha(E)} \exp(-iE/2)\} \\ \nonumber
\simeq -f(E)(1+3/4\log E+\cdots),
\end{eqnarray*} 
where the asymptotic limit applies provided that $U(\alpha,3/2,iE)\sim (iE)^{\alpha}$, when $E\gg 0$. Now  $f(E)=e^{iE/2}/F(\alpha,3/2,iE)$ (which indeed is  a real-valued function)  may also be formally  written as a Taylor series of $\log E$. The procedure obtains $\tan\delta_{0}$  as a series of  $\log E$ times $\sqrt{N}$, to be inverted near $E=1/3\log \sqrt{N}$  as:
 \[\delta_0\simeq \pi/2 +A(N)\sqrt{N}\log E- h(N)\]  
 for some  $A(N)$ and $h(N)$ (two parameters being the sum of the series computed at that point). This method is valid  whenever $k\gg 1$, i.e. $x\ll \sqrt{N}$.

\section{Asymptote $\pi(x|N)$ for $x\ll \sqrt{N}$.}

\subsection{Calculation of $u(\kappa)$.}
For finite $N$, a interpolating polynomial $u(\kappa)$ of degree two in $\kappa$ and free term equal to zero is  interpolated (Lagrange conditions): 

\begin{eqnarray}
\label{eq:u_parabolic}
u(N,x)\simeq \alpha_{1}(N)\kappa -\alpha_{2}(N)\kappa^2
\end{eqnarray}

The parameters $\alpha_1$ and $\alpha_2$  are chosen to fit the values at two different primes, $x=p_1$ and $x=p_2$   (note that $u$ is a step function when $\kappa\sim O(1)$. See  Fig.~\ref{ukappa}  obtained  for $\mathcal{F}(304)$),  i.e.: 

\begin{eqnarray*}
u(1-  \frac{\sqrt{N} }{|\mathcal{F}|}  \sum_{p=2}^{p_2} \frac{1}{p}) \simeq  \gamma \log(\sqrt{N}/p_2), \\ 
u(1-\frac{\sqrt{N} }{|\mathcal{F}|}  \sum_{p=2}^{p'_1} \frac{1}{p})\simeq \gamma \log(\sqrt{N}/p_1),
\end{eqnarray*}
where $p'_1$ is the prime number before $p_1$.

\begin{figure}[hbtp]
\centering
\includegraphics[scale=0.6]{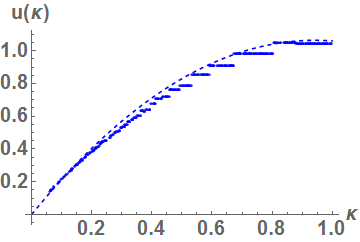}
\caption{ The function $u$ versus $\kappa=k/|\mathcal{F}|$ for $x$ in the ensemble of factorization with $j=304$. Note that the quadratic approximation, although computed for small $x$ and large $k=\kappa |\mathcal{F}|$, fits perfectly well even for large $x$ and small $k$. }
\label{ukappa}
\end{figure}

From these we get in the asymptotic (large $N$) regime: 

  \[
  \alpha_2 \simeq (1-\tau)/(1-\nu),
  \] 
  
  \[
  \alpha_1 \simeq 1+\alpha_2+\gamma\log \frac{p_2}{p_1}.
  \] 
  with
   \[\nu = \sum_{p=p_1}^{p_2} \frac{1}{p} \sqrt{N}/|\mathcal{F}| ,\]
 
       \[\tau=\frac{\gamma}{\nu}(1-\nu)\log\frac{p_2}{p_1}\] and \[|\mathcal{F}|\sim \sqrt{N}\log\log \sqrt{N}.\]  
  
  In the main text, we have chosen $p_1=2$ and $p_2=3$. We have tested for many different values without finding any significant difference.
  
  This solution, that is (in principle, given the high degree of accuracy that can be seen in Fig.3) valid for any $N$ can be also used as a further proof of the quantum simulator. It cannot be by chance that this level of accuracy is obtained. We can go further and check explicitly that the statistics of the states corresponds to that of the primes.

\subsection{Calculation of $E(x)$.}
\begin{figure}[hbtp]
\centering
\includegraphics[scale=0.35]{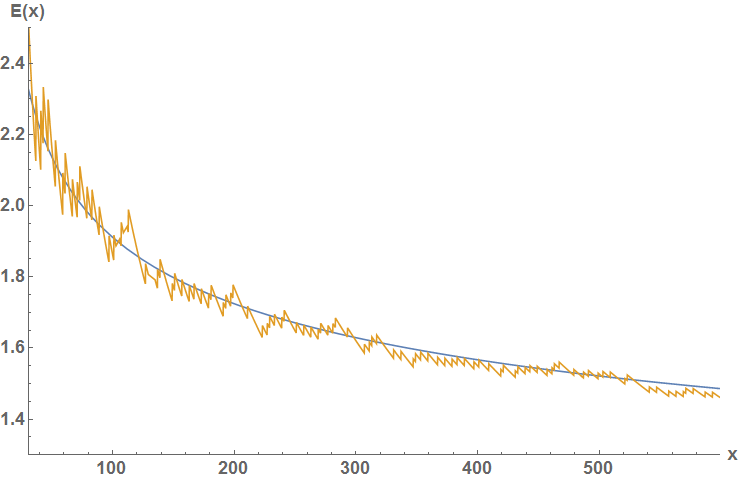}
\caption{ The spectrum of the simulator: the actual values of $\pi(x)\pi(N/x)/j^2$ for $N$'s (orange) in the factorization ensemble with $j=90004$ represented together with the quantum formula Eq.~\ref{eq:egamma} (blue). } 
\label{Ex}
\end{figure}

Inverting Eq. \ref{eq:u_parabolic}, provides the spectrum of $E_k$ from semi-classical quantization:
\begin{equation}
\label{eq:egamma}
E\rightarrow C\gamma^{-\kappa(x)}\simeq C\gamma^{ (\alpha_1^2/(2\alpha_2)^2-  u(x)/\alpha_2)^{1/2} -\alpha_1/(2\alpha_2)}.
\end{equation}
$C$ is just a parameter of $N$. The correctness of Eq. \ref{eq:egamma} can be seen in Fig. \ref{Ex}. This spectrum can be seen as a prediction of the simulator.

\subsection{Formal derivation of the asymptotic behavior of $\pi(x|N)$.} Formally, when $x|N$,

\begin{equation}
\label{eq:pifromE}
\pi(x)=j^2 E(x)/\pi(N/x)
\end{equation}
Note that we are concerned about the semiclassical theory of the simulator for those $x\ll \sqrt{N}$, i.e., equivalently we are considering values of $\pi(N/x)\gg \pi(\sqrt{N})=j\gg 1$. Then, we are allowed to use the prime number theorem to derive the form of $\pi(N/x)$ for $N/x\gg 1$.

First, from the definition of $u(N,x)$, $N/x=\sqrt{N}\exp\{u/\gamma\}$ and $x=\sqrt{N}\exp\{-u/\gamma\}$,

\[ \pi(N/x)\rightarrow \frac{\frac{N}{x}}{\log N- \log x}=\frac{\sqrt{N}\exp\{u/\gamma\}}{\log \sqrt{N}(1+\frac{u}{j}\frac{\sqrt{N}}{\log\sqrt{N}})},\]
but, $j\rightarrow \frac{\sqrt{N}}{\log\sqrt{N}}$, and since $\exp\{u/\gamma\}=\sqrt{N}/x$, this gives

\begin{equation}
\pi(N/x)\rightarrow \frac{j}{1+u}\frac{\sqrt{N}}{x}.
\end{equation}

This result, when $x=3$, can be used to obtain the maximum of $E$ in the factorization ensemble (Eq. 17 in the main text).

Feeding this in Eq.~\ref{eq:pifromE}, obtains the asymptote 

\begin{equation}
\label{eq:piapprox}
\pi(x|N)\rightarrow \gamma x\;(1+u)E(x).
\end{equation}

Eq. \ref{eq:piapprox} allows also to compute the constant $C$: If we know the value of $\pi(x_0)$ for $x_0\ll \sqrt{N}$ then:

\begin{equation}
\label{eq:C}
C= \frac{\pi(x_0)}{x_0(1+u(N, x_0))}\gamma^{\kappa(x_0)-1};
\end{equation}
Particularly, for $N\rightarrow \infty$, we can safely take $x_0=3$, $\pi(x_0)=2$,  $\kappa(x_0)\rightarrow 1$ and $u(N,x_0)\rightarrow 1$, obtaining $C\rightarrow 1/3$, independently of $N$ as it should be.  

\begin{figure}[hbtp]
\centering
\includegraphics[scale=0.35]{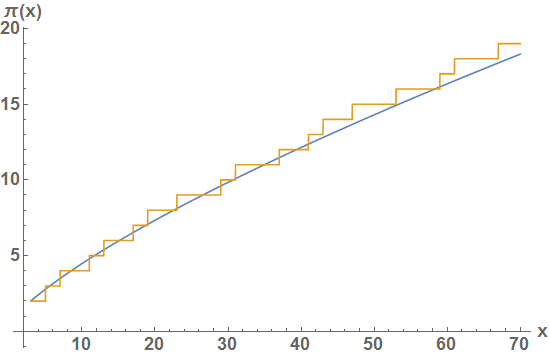}
\caption{ The functions $\pi(x|N)$ (blue) and $\pi(x)$ (orange) for $x$ in the ensemble of factorization with $j=139$, ($N=661643$). }
\label{pix661643}
\end{figure}

\begin{figure}[hbtp]
\centering
\includegraphics[scale=0.35]{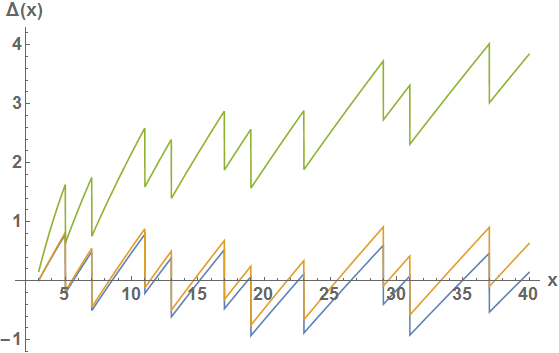}
\caption{ The functions $\pi(x|N)-\pi(x)$ (blue), $R(x)-\pi(x)$ (orange) and  $Li(x)-\pi(x)$ (green) for $x$ in the ensemble of factorization with $j=139$, ($N=661643$). Note how $\pi(x|N)$ fits perfectly to the best  analytical approximation given by $R(x)$ for $x\ll \sqrt{N}$.} 
\label{deltax661643}
\end{figure}

\begin{figure}[hbtp]
\centering
\includegraphics[scale=0.25]{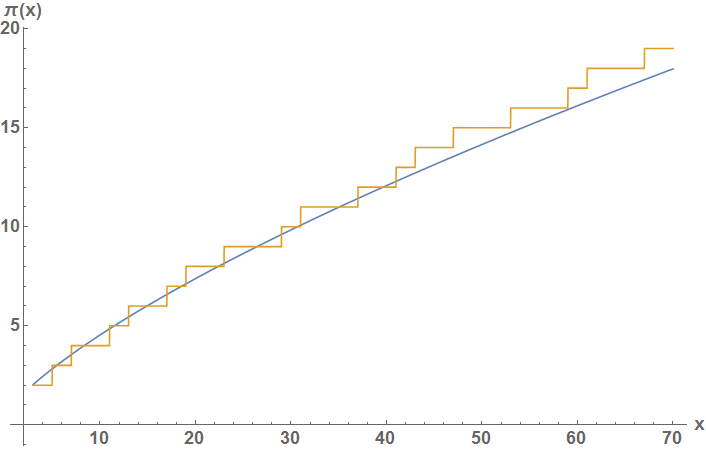}
\caption{ The functions $\pi(x|N)$ (blue) and $\pi(x)$ (orange) for $x$ in the ensemble of factorization with $j=62$, ( $N=91301$). }
\label{pix91301}
\end{figure}

\begin{figure}[hbtp]
\centering
\includegraphics[scale=0.4]{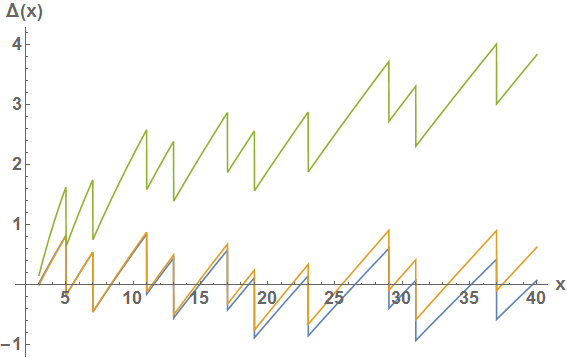}
\caption{ The functions $\pi(x|N)-\pi(x)$ (blue),  $R(x)-\pi(x)$ (orange) and  $Li(x)-\pi(x)$ (green) for $x$ in the ensemble of factorization with $j=62$, ($N=91301$). Note how $\pi(x|N)$ fits perfectly to the best  analytical approximation given by $R(x)$ for $x\ll \sqrt{N}$.} 
\label{deltax91301}
\end{figure}

\begin{figure}[hbtp]
\centering
\includegraphics[scale=0.35]{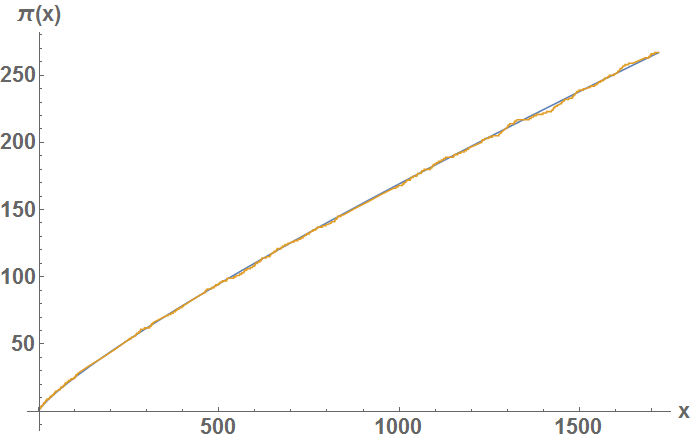}
\caption{ The functions $\pi(x|N)$ (blue) and $\pi(x)$ (orange) for $x$ in the ensemble of factorization with $j=1000000$, ($N=239811952854769$). }
\label{bigN1}
\end{figure}

\begin{figure}[hbtp]
\centering
\includegraphics[scale=0.30]{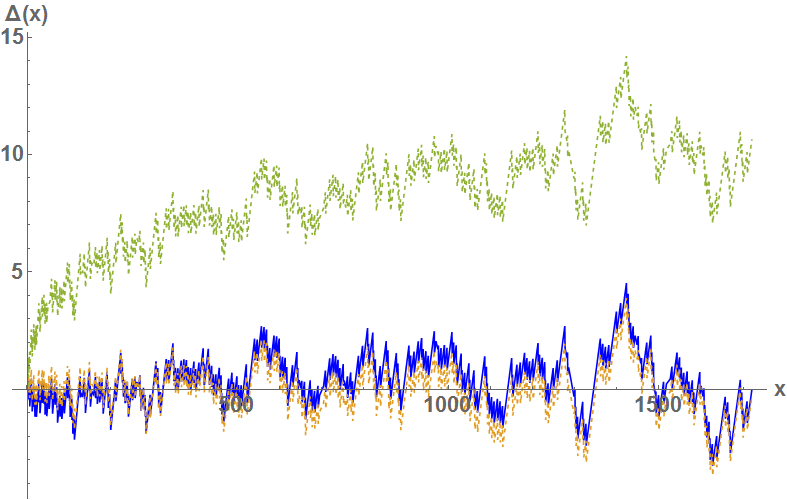}
\caption{ The functions $\pi(x|N)-\pi(x)$ (blue), $R(x)-\pi(x)$ (orange) and  $Li(x)-\pi(x)$ (green) for $x$ in the ensemble of factorization with $j=1000000$, ($N= 15485863^2=239811952854769$). Note how $\pi(x|N)$ fits perfectly to the best  analytical approximation given by $R(x)$ for $x\ll \sqrt{N}$.} 
\label{bigN2}
\end{figure}

In order to show the precision of the new expression for $\pi(x)$, we can calculate the graphs of $\pi(x|N)$ for other values of $N$ and plot them together. This is done  in Fig.\ref{pix661643} and Fig.\ref{pix91301} where $\pi(x|N)$ is displayed together with the exact $\pi(x)$ for the values $N=661643$ and $N=91301$. To further demonstrate its precision we  can also  compare this with the best possible approximation, given by Riemann for $x\ll \sqrt{N}$. This is shown in Figures \ref{deltax661643} and \ref{deltax91301}. It is to be noted that this is  a result of the exact solution, thus allowing us to be very confident in the relevance of the results. Note also that the new expression is much better that $Li(x)$.   In Fig. \ref{bigN1} and Fig. \ref{bigN2} we see the result of $\pi(x/N)$ and $\pi(x|N)-\pi(x)$ for $j=1000000$.



\end{document}